\documentclass[12pt]{article}
\begin{document}
\title{\huge \bf Octonionic Version of Dirac Equations}
\author{{\bf Merab Gogberashvili} \\
Andronikashvili Institute of Physics \\
6 Tamarashvili St., Tbilisi 0177, Georgia\\
{\sl E-mail: gogber@hotmail.com }}
\maketitle
\begin{abstract}
It is shown that a simple continuity condition in the algebra of
split octonions suffices to formulate a system of differential
equations that are equivalent to the standard Dirac equations. In
our approach the particle mass and electro-magnetic potentials are
part of an octonionic gradient function together with the
space-time derivatives. As distinct from previous attempts to
translate the Dirac equations into different number systems here
the wave functions are real split octonions and not bi-spinors. To
formulate positively defined probability amplitudes four different
split octonions (transforming into each other by discrete
transformations) are necessary, rather then two complex wave
functions which correspond to particles and antiparticles in usual
Dirac theory.
\end{abstract}
\medskip


\section{Introduction}

One of the breakthroughs in the development of field theory was
the discovery of the Dirac equation in 1928. The question we wish
to address in this article is whether one can formulate the Dirac
equation without availing oneself of complex bi-spinors and matrix
algebra, and to what extent such a formulation can be brought into
a form equivalent to the standard theory. The successful
application of quaternions \cite{QuatD} and Geometric Algebras
\cite{CliffD} in formulating a Dirac equation without matrices,
initiated the use of octonions as underlying numerical fields
\cite{OctD,LeAb}. Real octonions also contain eight parameters,
just as Dirac bi-spinors, or complex quaternions.

Octonions form the widest normed algebra after the algebras of
real numbers, complex numbers, and quaternions \cite{Sc}. Since
their discovery in 1844-1845 by Graves and Cayley there have been
various attempts to find appropriate uses for octonions in physics
(see reviews \cite{Oct}). One can point to the possible impact of
octonions on: Color symmetry \cite{Color}; GUTs \cite{GUT};
Representation of Clifford algebras \cite{Cliff}; Quantum
mechanics \cite{QM}; Space-time symmetries \cite{Rel}; Field
theory \cite{QFT,Adl}; Formulations of wave equations
\cite{OctD,LeAb,WE}; Quantum Hall effect \cite{Hall}; Strings and
$M$-theory \cite{String}; {\it etc}.

The structure of the matrices in the Dirac equation is linked to
relativistic covariance. However, space-time geometry does not only
have to be formulated using standard Lorentz four-vectors, matrices, or
quaternionic notation, but can also be formulated using an octonionic
parametrization.

In our previous papers \cite{Go} it was introduced the concept of
Octonionic Geometry based on the algebra of split octonions. This
approach is related with so-called Geometric Algebras \cite{GeoA}
in the sense that we also emphasized the geometric significance of
vectors (which are more effective than spinors and tensors in
conveying geometry) and avoided matrices and tensors.

In the present paper we shall show that the algebra of split
octonions, we used in \cite{Go} to describe the geometry, suffice
to formulate a system of differential equations equivalent to the
standard Dirac equations.


\section{Octonionic Geometry}

The geometry of space-time in the language of algebras and
symmetries can be described. Any observable quantity, which our
brain could extract from a single measurement is a real number.
Introduction of the distance (norm) always means some comparison
of two physical objects using one of them as an etalon. In the
algebraic language these features mean that to perceive the real
world our brain uses normed algebras with the unit element over
the field of real numbers. In physical applications of normed
algebras mainly the elements with the negative square (which are
similar to ordinary complex unit) are used. In this case norm of
the algebra is positively defined. Introduction of vector-like
elements with positive square and negative norm leads to so-called
split algebras. Because of pseudo-Euclidean character of there
norms split-algebras are useful to study dynamics.

In the paper \cite{Go} it was assumed that to describe the
geometry of real world most convenient is the algebra of
split-octonions. With a real physical signal we associate an
8-dimensional number, the element of split octonions,
\begin{equation} \label{s}
s = ct  + x^nJ_n + \hbar \lambda^nj_n + c\hbar\omega I~. ~~~~~(n =
1, 2, 3)
\end{equation}
Some characteristics of the physical world (such as dimension,
causality, maximal velocities, quantum behavior, {\it etc.}) can
be naturally connected with the structure of the algebra. For
example, our imagination about 3-dimensional character of the
space can be the result of existing of the three vector like
elements $J_n$ in (\ref{s}).

We interpret the basis elements of split octonions as
multi-vectors, similar to Geometric Algebras \cite{GeoA}. In
(\ref{s}) the scalar unit is denoted as $1$, the three vector-like
objects as $J_n$, the three pseudo-vectors as $j_n$ and the
pseudo-scalar as $I$. The eight scalar parameters that multiply
the basis units in (\ref{s}) we treat as the time $t$, the special
coordinates $x^n$, the wavelength $\lambda^n$ and the frequency
$\omega$. The quantity (\ref{s}) also contains two fundamental
constants of physics - the velocity of light $c$ and Planck's
constant $\hbar$. The appearance of these constants is connected
with the existence of two classes of zero divisors in the algebra
of split octonions \cite{Go}.

The algebra of the basis elements of split octonions can be
written in the form:
\begin{eqnarray} \label{algebra}
J_n^2 = - j_n^2 = I^2 = 1~, \nonumber \\
J_nj_m = - j_mJ_n = - \epsilon_{nmk}J^k~,  \nonumber\\
J_nJ_m = - J_mJ_n = j_nj_m = -j_mj_n = \epsilon_{nmk} j^k~, \\
J_nI = - IJ_n = j_n~, \nonumber \\
j_nI = -Ij_n = J_n~, \nonumber
\end{eqnarray}
where $\epsilon_{nmk}$ is the fully antisymmetric tensor and
$n,m,k = 1,2,3$. From these formulae it can be seen that to
generate a complete 8-dimensional basis of split octonions the
multiplication and distribution laws of only three vector-like
elements, $J_n$, are enough. The other two basis units $j_n$ and
$I$ can be expressed as binary and triple products
\begin{equation} \label{jI}
j_n = \frac{1}{2} \epsilon_{nmk}J^mJ^k~, ~~~~~I = J_nj_n
\end{equation}
(there is no summing of indices in the second formula).

The essential property of octonions, non-associativity, is the
direct result of the second formula of (\ref{jI}). Since the
3-vector $I$ has three equivalent representations we find, for
example,
\begin{equation}
J_1(J_2J_3) - (J_1J_2)J_3 = J_1j_1 - j_3J_3 = 2I \ne 0~.
\end{equation}
We adopt the non-associativity of the triple products of $J_n$,
and at the same time we need to have definite results for the
multiplication of all seven octonionic basis units
(\ref{algebra}). For this purpose the property of alternativity of
octonions can be used. This weak form of associativity implies the
Moufang identities for the products of any four element when two
of them coincide \cite{Sc}
\begin{equation} \label{moufang}
(ax)(ya) = a(xy)a~, ~~~~~ a(x(ay)) = (axa)y~, ~~~~~ y(a(xa)) =
y(axa) ~.
\end{equation}

In physical applications we interpret the non-associativity of
octonions, which results in the non-equivalence of left and right
products for expressions containing more than two basis units
$J_n$, as corresponding to causality \cite{Go}. For the direction
from the past to the future we want to use one definite order of
multiplication, for example the left product. Then
non-associativity leads to the appearance of time asymmetries in
our model.

The standard conjugation of fundamental vector-like basis units
\begin{equation}
J_n^* = - J_n~,
\end{equation}
can be imagined as reflections. Analogous to Clifford algebras
\cite{CliffD} we introduce three different kind of conjugations
(involutions) of products of several $J_n$ and thus define
conjugations of other octonionic basis units $j_i$ and $I$.
The standard octonionic anti-automorphism
\begin{equation} \label{*}
(J_iJ_kJ_n ...)^* = ...J_n^*J_k^*J_i^* ~,
\end{equation}
reverses the order of elements in any given expression.

We can define also the following automorphism (conjugation without
reversion)
\begin{equation}\label{dag}
(J_iJ_kJ_n ...)^\dag = J_i^* J_k^*J_n^* ... ~,
\end{equation}
and combined involution
\begin{equation}\label{overline}
\overline{(J_iJ_kJ_n ...)} = (J_iJ_kJ_n ...)^{\dag *} ~.
\end{equation}
Similar involutions in Clifford algebras are called respectively
conjugation, grade involution and reversion \cite{CliffD}.

Since we interpret the conjugation of vector-like elements $J_n$
as reflection and the directivity feature of their products as
corresponding to the time arrow, these three involutions
(\ref{*}), (\ref{dag}) and (\ref{overline}) can be considered
analogously to the discrete symmetries $T$, $P$ and $C$.

The involutions (\ref{*}), (\ref{dag}) and (\ref{overline}) do
not affect the unit elements of split octonions, while other basis
elements change according to the laws
\begin{eqnarray}
J_i^* = - J_i~, ~~~~~j_i^* = -j_i~, ~~~~~ I^* = -I~, \nonumber \\
J_i^\dag = - J_i~, ~~~~~j_i^\dag = j_i~, ~~~~~ I^\dag = -I~, \\
\overline{J_i} = J_i~, ~~~~~ \overline{j_i} = - j_i~, ~~~~~
\overline{I} = I~. \nonumber
\end{eqnarray}

The principal conjugation of octonions (\ref{*}) is usually used
to define their norm. For example, conjugation of  (\ref{s}) gives
\begin{equation} \label{s*}
s^* = ct - x_nJ^n - \hbar\lambda_nj^n - c\hbar\omega I~.
\end{equation}
Then the norm of (\ref{s})
\begin{equation} \label{sN}
s^2 = ss^* = s^*s = c^2t^2 - x_nx^n  + \hbar^2 \lambda_n\lambda^n
- c^2\hbar^2\omega^2 ~,
\end{equation}
has a $(4+4)$ signature and in the limit $\hbar \rightarrow 0$ gives
the classical formula for the Minkowski interval.

It is possible to give a representation of the basis units of
octonions ($1, J_n, j_n, I$) through $2 \times 2$ Zorn matrices
\cite{Sc}, whose diagonal elements are scalars and whose
off-diagonal elements are 3-dimensional vectors
\begin{eqnarray} \label{JjImatrix}
\begin{array}{cc}
1~\Longleftrightarrow \pmatrix{1 & 0 \cr 0 & 1 }, &
J_n\Longleftrightarrow \pmatrix{0 & \sigma_n \cr \sigma_n & 0},
\cr I~\Longleftrightarrow \pmatrix{1 & 0 \cr 0 & -1 }, &
j_n\Longleftrightarrow \pmatrix{0 & -\sigma_n \cr \sigma_n & 0}.
\end{array}
\end{eqnarray}
Here the elements  $\sigma_n$ ($n = 1, 2, 3$), with the property
$\sigma^2_n = 1$ can be considered as ordinary Pauli matrices. Of
course one can use the real representation also, with the complex
Pauli matrix $\sigma_2$ replaced by the unit matrix.

Using (\ref{JjImatrix}) the split octonion (\ref{s}) can be
written as
\begin{equation} \label{Omatrix}
s = ct + x_nJ^n + \hbar\lambda_nj^n + c\hbar\omega I =
\pmatrix{c(t+ \hbar\omega ) & (x_n-\hbar\lambda_n)\sigma^n \cr
(x_n+\hbar\lambda_n)\sigma^n & c(t-\hbar\omega)}~.
\end{equation}
The conjugate of the matrix (\ref{Omatrix}) is defined as
\begin{equation} \label{Omatrix*}
s^* = ct - x_nJ^n - \hbar\lambda_nj^n - c\hbar\omega I =
\pmatrix{c(t-\hbar\omega ) & -(x_n-\hbar\lambda_n)\sigma^n \cr
-(x_n+\hbar\lambda_n)\sigma^n & c(t+\hbar\omega )}~.
\end{equation}
Then the norm (\ref{sN}) is given by the product of these matrices
\begin{equation}
s^*s = (c^2t^2 - x_nx^n + \hbar^2\lambda_n\lambda^n -
c^2\hbar^2\omega^2)\pmatrix{1 & 0 \cr 0 & 1 } ~.
\end{equation}

The two other forms of involution, (\ref{dag}) and
(\ref{overline}), for the octonion (\ref{s}) have the following
matrix representations
\begin{eqnarray} \label{Omatrix+-}
s^\dag = ct - x_nJ^n + \hbar\lambda_nj^n - c\hbar\omega I =
\pmatrix{c(t - \hbar\omega ) & -(x_n + \hbar\lambda_n)\sigma^n \cr
-(x_n - \hbar\lambda_n)\sigma^n & c(t + \hbar\omega )}~, \nonumber \\
\overline{s} = ct + x_nJ^n - \hbar\lambda_nj^n + c\hbar\omega I =
\pmatrix{c(t + \hbar\omega ) & (x_n + \hbar\lambda_n)\sigma^n \cr
(x_n - \hbar\lambda_n)\sigma^n & c(t - \hbar\omega )}~.
\end{eqnarray}

Since octonions are not associative, they cannot be represented by
matrices with the usual multiplication laws. The product of any
matrices written above, have the special multiplication rules
\cite{zorn}
\begin{equation}\label{Zorn*}
\pmatrix{\alpha & {\bf a} \cr {\bf b} & \beta} * \pmatrix{\alpha '
& {\bf a'} \cr {\bf b'} & \beta' } =  \pmatrix{\alpha\alpha' +
({\bf a b'}) & \alpha {\bf a'} + \beta' {\bf a} - [{\bf bb'}] \cr
\alpha'{\bf b} + \beta {\bf b'} + [{\bf a a'}] & \beta\beta' +
({\bf b a'})},\nonumber
\end{equation}
where (${\bf ab}$) and [${\bf ab}$] denote the usual scalar and
vector products of the 3-dimensional vectors ${\bf a}$ and ${\bf
b}$. Probably the easiest way to think of this multiplication is to
consider the usual matrix product with an added anti-diagonal matrix
\begin{equation}
\pmatrix{0 & - \epsilon^{nmk}b_mb'_k \cr
\epsilon^{nmk}a_m a'_k & 0} ~.
\end{equation}
The algebra of octonionic basis units (\ref{algebra}) is easily
reproduced in this Zorn matrix notation.

Using the algebra of the basis elements (\ref{algebra}) the octonion
(\ref{s}) also can be written in the equivalent form
\begin{equation} \label{s'}
s = c(t + \hbar \omega I) +  J^n(x_n + \hbar \lambda_n I) ~.
\end{equation}
From this formula we see that pseudo-scalar $I$ introduces the
'quantum' term corresponding to some kind of uncertainty of
the space-time coordinates.


\section{Octonionic Dirac Equation}

For convenience we take the formulation of ordinary Dirac equation
in the notation used in \cite{Par}
\begin{equation} \label{dirac}
\left(i\hbar \gamma^0 \partial_0 + i\hbar \gamma^n
\partial_n - \frac{e}{c}\gamma^0A_0 +
\frac{e}{c}\gamma^nA_n - mc\right) \Psi = 0~,
\end{equation}
which as distinct from the standard definition \cite{BjDr} has the
opposite sign for $\gamma^0$ and scalar potential $A_0$. In
(\ref{dirac}) gamma matrices have the representation
\begin{equation} \label{gamma}
\gamma_0 = - \pmatrix{1 & 0 \cr 0 & -1 }~, ~~~~~ \gamma_n =
\pmatrix{0 & \sigma_n \cr \sigma_n & 0}~, ~~~~~ \gamma_5 =
i\pmatrix{0 & 1 \cr 1 & 0 }~,
\end{equation}
where $\sigma_n$ ($n = 1, 2, 3$) are the usual Pauli matrices.

Without any bias about the nature of the quantities involved in the
wave-function, including any hidden meaning to the imaginary unit
$i$, the standard  4-dimensional spinor can be written in the form
\begin{equation} \label{Psi}
\Psi = \left( \begin{array}{c}  y_0 + i L_3 \\
                               -L_2 + i L_1 \\
                                y_3 + i L_0 \\
                                y_1 + i y_2
\end{array} \right) ~.
\end{equation}
It is characterized by the eight real parameters $y^\nu$ and
$L^\nu$ ($\nu = 0, 1, 2, 3$).

The Dirac equation (\ref{dirac}) for the wave-function (\ref{Psi})
can be decomposed into an equivalent set of eight real
differential equations
\begin{eqnarray} \label{system}
-\frac{e}{c}A^\nu y_\nu  + \hbar (F_{03} + f_{21}) = mcy_0 ~,  \nonumber\\
\frac{e}{c}A^\nu L_\nu + \hbar (F_{12} + f_{03}) = mcL_0 ~, \nonumber\\
\frac{e}{c}\left([yA]_{10} + [LA]_{32}\right) +
\hbar (F_{13} + f_{20}) = mcy_1 ~, \nonumber \\
\frac{e}{c}\left([yA]_{20} + [LA]_{13}\right) +
\hbar (F_{23} + f_{01}) = mcy_2 ~, \nonumber \\
\frac{e}{c}\left([yA]_{30} + [LA]_{21}\right) -
\hbar\partial^\nu L_\nu = mcy_3 ~, \\
\frac{e}{c}\left([yA]_{32} + [LA]_{01}\right) +
\hbar (F_{02} + f_{13}) = mcL_1 ~, \nonumber \\
\frac{e}{c}\left([yA]_{13} + [LA]_{02}\right) +
\hbar (F_{10} + f_{23}) = mcL_2 ~, \nonumber \\
\frac{e}{c}\left([yA]_{12} + [LA]_{03}\right) - \hbar\partial^\nu
y_\nu = mcL_3 ~, \nonumber
\end{eqnarray}
where the summing is done by the Minkowski metric $\eta^{\nu\mu}$
$(\nu, \mu = 0,1,2,3)$ with the signature $(+---)$ and we have
introduced the notations
\begin{eqnarray} \label{Ff}
f_{\nu\mu} = \partial_\nu y_\mu - \partial_\mu y_\nu ~, ~~~~
[LA]_{\nu\mu} = L_\nu A_\mu - L_\mu A_\nu ~,
 \nonumber\\
F_{\nu\mu} =
\partial_\nu L_\mu - \partial_\mu L_\nu ~, ~~~~
[yA]_{\nu\mu} = y_\nu A_\mu - y_\mu A_\nu~.
\end{eqnarray}

Now we want to show that the system (\ref{system}) can be written
as the product of two split octonions with real components.

The particle wave function we denote by the split octonion
\begin{equation} \label{psi}
\psi = - y_0 + y_nJ^n + L_nj^n + L_0 I ~, ~~~~~(n = 1,2,3)
\end{equation}
where we use the same notation for the eight real numbers $y^\nu$
and $L^\nu$ as for the components of the Dirac wave-function
(\ref{Psi}). The set of the values of $y^\nu$ and $L^\nu$ can be
understood as the traditional functions depending on the frame,
since after any measurement one finds new values for these
parameters that depend on the observer's frame.

To the measurement process we want to associate a split octonion
having the dimension of momentum
\begin{equation} \label{nabla}
\nabla = \hbar \left[c\frac{\partial}{\partial t} +
J^n\frac{\partial}{\partial x^n} \right]  + \left[-\left(mc +
\frac{e}{c}A_0\right) + \frac{e}{c} A_n J^n \right]I ~.
\end{equation}
In this gradient function we use standard notations for the
coordinates, mass and components of vector-potential. In the limit
$m, A_\nu \rightarrow 0$ the norm of $\nabla $ is the ordinary
d'Alembertian.

Now let us write the orientated continuity equation by multiplication
of the octonionic wave-function (\ref{psi}) by (\ref{nabla}) from
the left
\begin{equation} \label{continuity}
\nabla_L \psi = \hbar \left[c\frac{\partial}{\partial t} +
J^i\frac{\partial}{\partial x^i} \right] \psi + \left[ -\left(mc +
\frac{e}{c}A_0\right) + \frac{e}{c} A_n J^n\right](I\psi) = 0 ~.
\end{equation}
Because of non-associativity it is crucial in the second term to
multiply $I$ first with $\psi$ and then with the remaining terms
in the brackets. The orientated product (\ref{continuity}) is
similar to the barred operators considered in \cite{LeAb}.

Using the matrix representation of octonionic basis units
(\ref{JjImatrix}) the quantities entering the equation
(\ref{continuity}) can be written in the form
\begin{eqnarray}
\begin{array}{c}
\psi = \pmatrix{y_0+L_0 & (y_n-L_n)\sigma^n \cr
(y_n+L_n)\sigma^n  & y_0-L_0 }, \nonumber\\
(I\psi) = \pmatrix{y_0+L_0 & (y_n-L_n)\sigma^n \cr
-(y_n+L_n)\sigma^n  & -y_0+L_0 }, \\
\left[c~\partial /\partial t + J^n\partial /\partial x^n \right] =
\pmatrix{c~\partial /\partial t & \sigma^n \partial /\partial x^n
\cr
\sigma^n \partial /\partial x^n & c~\partial /\partial t },  \\
\left[ -\left(mc + A_0 e/c\right) + J^nA_n e/c\right] =
\pmatrix{-(mc + A_0e/c)  & \sigma^n A_n e/c \cr \sigma^n A_n e/c &
-(mc + A_0e/c)  }. \nonumber
\end{array}
\end{eqnarray}

According to the multiplication rules of octonionic matrices
(\ref{Zorn*}), or using the algebra of the basis units
(\ref{algebra}), equation (\ref{continuity}) takes the form:
\begin{eqnarray} \label{nablaPsi}
\left[-mcy_0 -\frac{e}{c}A^\nu y_\nu  + \hbar\partial^\nu L_\nu
\right] + \left[-mcL_0-\frac{e}{c}A^\nu L_\nu  +
\hbar\partial^\nu y_\nu \right]I + ~~~~~ \nonumber \\
+ \left[ -mcy^i +\frac{e}{c}[Ay]^{i0} + \hbar F^{i0} +
\epsilon^{ijk} \left(\hbar f_{jk} +
\frac{e}{c}A_jL_k\right)\right]J_i + ~~~~~ \\
+ \left[-mcL^i + \frac{e}{c}[AL]^{i0} + \hbar f^{i0} -
\epsilon^{ijk} \left(\hbar F_{jk} +
\frac{e}{c}A_jy_k\right)\right]j_i = 0 ~, \nonumber
\end{eqnarray}
where $\nu = 0,1,2,3$ and $i, j, k = 1, 2, 3$.

Now we have all the tools to reproduce the Dirac equations.
Equating to zero coefficients in front of octonionic basis units
in (\ref{nablaPsi}) we have a system of eight equations.
Subtracting the pairs of these equations by the rules $(J_3 -
j_3)$, $(j_2 - j_1)$, $(1 - J_2)$ and $(J_1 - I) $ we arrive at a
system of four 'complex' equations
\begin{eqnarray} \label{nablaPsi'}
\frac{e}{c}[ - A_0 (y_0 + iL_3) + A_1 (y_1 + i y_2)+
A_2(y_2 - iy_1) + A_3(y_3 + iL_0)] -\nonumber \\
- \hbar [\partial_0(L_3 - iy_0) - \partial_1(y_2 - iy_1) +
\partial_2(y_1 - iy_2) - \partial_3 (L_0 - iy_3)]  = cm (y_0 + iL_3) ,  \nonumber\\
\nonumber\\
\frac{e}{c}[- A_0 (-L_2 + iL_1) + A_1 (y_3 + i L_0) -
A_2(L_0 - iy_3) - A_3(y_1 + iy_2)] - \nonumber \\
- \hbar [\partial_0(L_1 + iL_2) - \partial_1(L_0 - iy_3) -
\partial_2(y_3 + iL_0) + \partial_3 (y_2 - iy_1)] = cm (iL_1 - L_2) ,  \nonumber\\
\\
\frac{e}{c}[ - A_0 (y_0 + iL_3) + A_1 (y_1 + i y_2) +
A_2(y_2 - iy_1) + A_3(y_3 + iL_0)] + \nonumber \\
+ \hbar [\partial_0(L_0 - iy_3) - \partial_1(L_1 + iL_2) -
\partial_2(L_2 + iL_1) - \partial_3 (L_3 + iy_0)] = cm (y_3 + iL_0) , \nonumber \\
\nonumber\\
\frac{e}{c}[ - A_0 (y_0 + iL_3) + A_1 (y_1 + i y_2) +
A_2(y_2 - iy_1) + A_3(y_3 + iL_0)] +  \nonumber \\
+ \hbar [\partial_0(y_2 - iy_1) - \partial_1(L_3 - iy_0) -
\partial_2(y_0 + iL_3) + \partial_3 (L_1 + iL_2)] = cm (y_1 +
iy_2) .\nonumber
\end{eqnarray}
Introducing the complex bi-spinor (\ref{Psi}) and gamma matrices
(\ref{gamma}) this system is easy to rewrite in matrix form, which
exactly coincides with the standard Dirac equations (\ref{dirac}).

To define the probability amplitudes we need a real modulus
function corresponding to a positively defined probability. It
cannot be the norm of the 'wave-function' (\ref{psi}), since the
norm of split octonions is not positively defined and thus cannot
give a satisfactory interpretation as a probability. However, the
product
\begin{equation} \label{N}
N = \frac{1}{2}\left[\psi \overline{\psi} + (\psi
\overline{\psi})^*\right] = \frac{1}{2}\left[\psi \overline{\psi}
+ \psi^\dag \psi^*\right]= y_0^2 + y_ny^n + L_nL^n + L_0^2~,
\end{equation}
can serve as the probability amplitude similar to standard Dirac
theory. The product (\ref{N}) is analogous to the Hermitian norm
introduced in \cite{DaDe}.


\section{Concluding Remarks}

In this paper it was shown that the Dirac equations can be written
using the algebra of split octonions over the field of real
numbers. As distinct from the similar models we do not use any
complex numbers or bi-spinors.

We wish to stress that because of non-associativity our model is
not equivalent to the standard Dirac theory. Without fixing the
order of multiplications the products of wave-functions,
corresponding to some physical process, will give not
single-valued results. This property, when left and right products
are not equivalent, in physical applications is natural to connect
with the time asymmetry and causality.

Differences with standard theory arise also when we try to
determine what kind of number should be used for the probability
amplitudes. To formulate positively defined probability amplitudes
four different split octonions (transforming into each other by
discrete transformations) are necessary, rather then two complex
wave functions corresponding to particles and antiparticles in
usual Dirac theory. This means that in the octonionic model
standard classification scheme of particles should be revised.

Note that the Hermitian norm (\ref{N}), which we want to associate
with the probability amplitude, exhibits more symmetry than the
standard norm of split octonions (having $(4+4)$ signature). It is
known that the automorphism group of ordinary octonions is the
smallest exceptional Lie group - $G_2$ \cite{Sc}, while the
automorphism group of split octonions is the real, non-compact
form of $G_2$. Some general results about the real, non-compact
$G_2$ and its subgroup structure can be found in \cite{BHW}.


\section*{Acknowledgments}

I want to acknowledge the hospitality of the Departments of
Physics of California State University (Fresno) and the University
of Minnesota where this paper was prepared.

The work was supported in part by a 2003 COBASE grant and by DOE
grant DE-FG02-94ER408.


\end{document}